\documentstyle[12pt,aasms4,graphics,psfig]{article}
\newcommand{\etal}{et~al.~}
\newcommand{\ubvri}{\protect\hbox{$U\!BV\!RI$} }

\def\simlt{\mathrel{\hbox{\rlap{\hbox{\lower4pt\hbox{$\sim$}}}\hbox{$<$}}}}
\def\simgt{\mathrel{\hbox{\rlap{\hbox{\lower4pt\hbox{$\sim$}}}\hbox{$>$}}}}

\def\ale{\mathrel{\hbox{\rlap{\hbox{\lower4pt\hbox{$\sim$}}}\hbox{$<$}}}}
\def\age{\mathrel{\hbox{\rlap{\hbox{\lower4pt\hbox{$\sim$}}}\hbox{$>$}}}}
\def\farcs{\hbox{$.\!\!^{\prime\prime}$}}
\def\farcm{\hbox{$.\mkern-4mu^\prime$}}
\def\fs{\hbox{$.\!\!^{\rm s}$}}

\begin{document}
\oddsidemargin=0mm

\title{
Identification of the Red Supergiant Progenitor of Supernova 2005cs: 
Do the Progenitors of Type II-P Supernovae Have Low Mass?
}

\author{Weidong~Li$^{1}$,
  Schuyler~D.~Van~Dyk$^{2}$,
  Alexei~V.~Filippenko$^{1}$,
  Jean-Charles~Cuillandre$^{3}$,
  Saurabh~Jha$^{1}$,
  Joshua~S.~Bloom$^{1}$,
  Adam~G.~Riess$^{4}$, \&
  Mario~Livio$^{4}$
}

\altaffiltext{1}{ Department of Astronomy,
	University of California,
        Berkeley, CA 94720-3411; email (wli, alex, sjha, jbloom)@astro.berkeley.edu }

\altaffiltext{2} {Spitzer Science Center,
    California Institute of Technology,
    Mailcode 220-6, Pasadena, CA 91125; email vandyk@ipac.caltech.edu }

\altaffiltext{3}{Canada-France-Hawaii Telescope Corporation,
     65-1238 Mamalahoa Hwy,
     Kamuela, HI 96743; email jcc@cfht.hawaii.edu }

\altaffiltext{4}{Space Telescope Science Institute, 3700 San Martin Drive, Baltimore, MD 21218;
     email (ariess, mlivio)@stsci.edu }

\begin{abstract}
\phantom{}\vskip 0.05cm


The stars that end their lives as supernovae (SNe)
have been directly observed in only a handful of cases, due
mainly to the extreme difficulty in identifying them in images obtained
prior to the SN explosions. Here we report the identification of the
progenitor for the recent Type II-plateau (core-collapse) SN 2005cs in
pre-explosion archival images of the Whirlpool Galaxy (M51) obtained with the
{\it Hubble Space Telescope\/} ({\it HST}) Advanced Camera for Surveys
(ACS).  From high-quality ground-based images of the SN from the
Canada-France-Hawaii Telescope, we precisely determine the
position of the SN and are able to isolate the SN progenitor to within
$0{\farcs}04$ in the {\it HST\/} ACS optical images.  We further pinpoint the
SN location to within $0{\farcs}005$ from {\it HST\/} ACS ultraviolet
images of the SN, confirming our progenitor identification.
From photometry of the SN progenitor obtained with the
pre-SN ACS images, and also limits to its brightness in pre-SN
{\it HST\/} NICMOS images, we infer that the progenitor is a red
supergiant star of spectral type K0--M3, with initial mass
7--9~$M_\odot$.  We also discuss the implications of the SN 2005cs
progenitor identification and its mass estimate. There is an emerging 
trend that the most common Type II-plateau SNe originate from 
low-mass supergiants (8--15~$M_\odot$). 

\end{abstract}



\section{Introduction}

Direct identification of the progenitors of supernovae (SNe) provides vital
information on their explosion mechanisms, a key issue in SN
studies. The white dwarfs thought to give rise to thermonuclear Type Ia
SNe have such exceedingly low luminosities that they cannot at present
be detected in external galaxies.  Core-collapse SNe (of Type II, Type Ib, and
Ic), on the other hand, arise
from far more luminous, massive stars. Unfortunately, even these progenitors
are so faint that their detection (ground-based or space-based) is
confined to nearby galaxies (distances $\ale$10 Mpc), in which SN
discoveries are relatively rare.  

Until now, out of more than 3,200 SNe discovered
since 1885, only 5 genuine SNe have had their progenitors identified: SN
1987A (a peculiar, subluminous SN~II) in the Large Magellanic Cloud
(e.g., Gilmozzi \etal 1987; Sonneborn \etal 1987), SN 1993J (an unusual
SN~IIb) in NGC 3031 (M81; Aldering \etal 1994; Van Dyk et al. 2002),  
SN 1999ev (Type II) in NGC 
4274 (Maund \& Smartt 2005), SN 2003gd (Type II) in NGC 628
(M74; Van Dyk \etal 2003a; Smartt \etal 2004), and SN 2004et (Type II) in NGC
6946 (Li \etal 2005a, 2005b).  In addition, identifications of
precursors that give rise to the super-outbursts of luminous blue
variable stars in other galaxies, occasionally misclassified as SNe,
have been made for several other objects (Zwicky 1964; Ryder \etal 1993;
Van Dyk \etal 1999b; Van Dyk \etal 2000).

Nature provided us with a rare opportunity to increase the sample of
directly identified SN progenitors with the discovery of SN 2005cs in the
Whirlpool Galaxy (hereafter, M51). The supernova was discovered by
Kloehr (2005) at about  magnitude 14 on 2005 June 28.9
(UT times are used throughout this paper), and confirmed in images obtained
with the 0.76-m Katzman
Automatic Imaging Telescope (KAIT; Li \etal 2000; Filippenko \etal 2001) at Lick Observatory
on 2005 June 30.25 (Li 2005). An optical spectrum taken with
the F. L. Whipple Observatory 1.5-m Tillinghast telescope on
2005 June 30.23 showed SN
2005cs to be a young SN~II (Modjaz \etal 2005), with P-Cygni-like line
profiles of the hydrogen Balmer series and helium superimposed on a
blue continuum.

Just five months before the discovery of SN 2005cs, M51
was fortuitously observed by the Hubble Heritage team (GO/DD program 10452;
PI: S. Beckwith) with the Wide Field Channel (WFC) of the Advanced 
Camera for Surveys (ACS) on-board the {\it Hubble Space Telescope\/}
({\it HST}). A large
four-color (F435W, F555W, F658N, and F814W) mosaic image of the nearly
face-on spiral galaxy NGC 5194 (M51a, the SN 2005cs host) and its
interacting companion, NGC 5195 (M51b), was obtained in six ACS
pointings (with four dithered exposures at each pointing), and the resulting
color composite image was released to the community on 2005 April 25
to celebrate the fifteenth anniversary of the successful operation of {\it
HST\/} (Mutchler \etal 2005).  These high-resolution ($0{\farcs}05$ pixel$^{-1}$)
images are also the deepest ever obtained of M51, reaching limiting
magnitudes of 27.3, 26.5, and 25.8 in the combined F435W ($\sim B$),
F555W ($\sim V$), and F814W ($\sim I$) images, respectively.  M51 had
also been observed in several bands with the Wide Field and Planetary
Camera 2 (WPFC2), but not to this depth or, generally, at such high 
spatial resolution.

M51 was also observed by the Near Infrared Camera and Multi-Object
Spectrometer (NICMOS) on-board {\it HST} in GTO program 7237
(PI: N. Scoville) in Cycle 7; the SN site was imaged in
five bands (F110W, F160W, F187N, F190N, and F222M) on 1998 June 28.
Together, the ACS and NICMOS data therefore provide images of unprecedented 
quantity and quality for possibly identifying and studying the progenitor 
star of SN 2005cs.

In \S~2 we report on our direct identification of the progenitor
of SN 2005cs, and in \S~3 we describe the progenitor's nature inferred
from analysis of these pre-SN {\it HST\/} data.  Further discussion 
is in \S~4, and we summarize our conclusions in \S~5.
We note that M51 was also host to SN 1945A (in NGC 5195; Type I) and  
SN 1994I (in NGC 5194; Type Ic).

\section{Identification of the Progenitor}

In Table 1 we list the {\it HST\/} ACS/WFC and NICMOS data that we
analyzed here.  Many additional pre-SN observations of M51 exist,
obtained with other
instruments on-board {\it HST\/}, including WFPC2 images of SN 1994I.
However, the ACS data provide the deepest and the highest-resolution
optical images currently available of the galaxy prior to the SN, supplemented
by the deep NICMOS data in the near-infrared.

\subsection{Registration of the Ground-based Observations}

To initially locate the SN 2005cs progenitor in these images, we
utilized the KAIT SN confirmation observations.  We identified 
six to eight
stars in common between the ACS/WFC and KAIT images and measured their
pixel coordinates. Using the task IRAF\footnote{IRAF
(Image Reduction and Analysis Facility) is distributed by the National Optical
Astronomy Observatories, which are operated by the Association of Universities
for Research in Astronomy, Inc., under cooperative agreement with the National
Science Foundation.} /GEOMAP, we performed
a geometrical transformation between the two sets of coordinates and
were able to match them to $\ale 3.0$ ACS pixels root-mean-square (rms $\ale
0{\farcs}15$). [This rather large uncertainty arises from the
relatively poor spatial resolution in the KAIT images ($0{\farcs}8$
pixel$^{-1}$) and the mediocre seeing under which the images were
obtained ($\sim$2$\arcsec.$5 FWHM).]  To within this positional uncertainty,
we were able to
identify a single object at the SN site in the F814W image.  However, this
object is not detected in the images in any of the other ACS passbands (Li \etal 2005c).

To better isolate the putative progenitor star, we obtained
higher-resolution ($0{\farcs}187$ pixel$^{-1}$) SN images with the
3.6-m Canada-France-Hawaii Telescope (CFHT, + MegaCam) on 
2005 July 2.28, under good seeing conditions ($0{\farcs}7$ FWHM). We
obtained 3$\times$10 s and 2$\times$90 s exposures in the Sloan $i$
band.   (Only the images from one chip containing the SN site, out
of a total of 36 chips, were analyzed.)  As a result,
we are able to detect many more faint objects in the CFHT images than in the
KAIT images. We identified 20--30 stars (or compact star clusters)
around the SN position that are present in both the CFHT and ACS
images. The CFHT images were geometrically transformed to the ACS
images using IRAF/GEOMAP, with uncertainties of $\ale 0.6$--0.8 ACS
pixel, and the SN location transformed onto the ACS images is
consistent in all five CFHT exposures (to within this
uncertainty). Increasing the number of stars included in the
transformation does not appreciably reduce the uncertainty, indicating
that we have approached the limit of the transformation accuracy.
An independent astrometric tie conducted using 90 stars or compact
clusters in both sets of images, fit to a third-order polynomial
function with full cross-terms for the transformation, resulted in a
consistent localization with a similar precision.

The final
error-weighted mean pixel location for all measurements from the CFHT
images of the SN is $X=4177.85 \pm 0.8$ and $Y=3421.17 \pm 0.8$ in
the ACS mosaic image ``h\_m51\_i\_s05\_drz\_sci\.fits" (Table 1), 
which corresponds to $\alpha({\rm
J2000})=13^h29^m52{\fs}764 \pm 0{\fs}004$, $\delta({\rm
J2000})=+47^\circ10\arcmin36{\farcs}09 \pm 0{\farcs}04$ from the ACS
image world coordinate system (WCS). Direct inspection of the F814W
image confirmed our initial identification of the candidate progenitor: only
one object is within the 0.8 ACS pixel uncertainty, at
position  $\alpha({\rm J2000})=13^h29^m52{\fs}760$,
$\delta({\rm J2000})=+47^\circ10\arcmin36{\farcs}11$ (Li \etal 2005d),
and we consider this object most likely to be the progenitor star.
The difference
between the transformed and measured position for this star is
$\Delta X = 0.71$ ACS pixel ($0{\farcs}036$) and $\Delta Y = 0.43$ ACS
pixel ($0{\farcs}022$), both of which are within 1$\sigma$ of the
positional uncertainties.

A comparison between the ACS/WFC color composite mosaic image and the
combined CFHT image is shown in Figure 1.

\subsection{Registration of the {\it HST}/ACS SN Observations}

Subsequent images of M51 showing SN 2005cs were obtained with {\it HST}/ACS
on 2005 July 11, and with NICMOS on 2005 July 13, as part of program
GO-10182 (PI: A.~V.~Filippenko).
The details of the observations are listed in Table 2.
The ACS images were observed with the High Resolution Channel (HRC) of
ACS. The HRC provides a finer spatial resolution
($0\farcs025$ pixel$^{-1}$) than the WFC ($0\farcs05$ pixel$^{-1}$),
but with a smaller field-of-view, $29\arcsec \times 25 \arcsec$
(compared to $202\arcsec \times 202\arcsec$ for WFC). 

Using six point sources within 10$\arcsec$ of the SN site, we
performed a geometrical transformation between the F250W
ACS/HRC SN image and the ACS/WFC F814W pre-SN image.
The transformation has an overall error of 0.1 ACS pixel ($0\farcs005$).
When the SN position measured from the ACS/HRC image is
transformed to the ACS/WFC image, it is coincident with our
progenitor candidate, with a difference of
$\Delta X = 0.12$ ACS pixel ($0{\farcs}006$) and $\Delta Y = 0.10$ ACS
pixel ($0{\farcs}005$). 
In Figure 2 we show a comparison between the ACS/HRC F250W
image and the ACS/WFC F814W image after image registration,
while in Figure 3 we show the SN site in all the four ACS/WFC
passbands. We now conclude to a high degree of certainty
that we have identified in the F814W image the progenitor of SN 2005cs.

\subsection{Registration of the {\it HST}/NICMOS Observations}

To determine whether the progenitor star is also seen in the pre-SN
{\it HST}/NICMOS observations, we attempted to match geometrically the
NICMOS images showing SN 2005cs to the pre-SN NICMOS
images. Out of the five NICMOS passbands in which M51 was imaged,
we consider only the F110W ($\sim J$) and F160W ($\sim H$) exposures
to be deep enough to be useful for our purpose.

We first identified five bright sources in common between
the new and the pre-SN NICMOS images.  We performed a
first-order geometrical transformation between the images which
achieved an rms accuracy of 0.08 NICMOS pixel
($0\farcs016$).
After the images were registered in common, we identified an additional 8--10
fainter objects in both sets of images.  A broader geometrical transformation then
resulted in a registration with rms accuracy 0.15 NICMOS pixel ($0\farcs03$).
This larger uncertainty in the transformation for the larger number of
fiducials is
due mainly to the measurement uncertainty in the centroids for
stars in the undersampled and relatively shallow NICMOS images.

In Figure 4 we show a comparison of the registered NICMOS
F110W images before and after SN 2005cs. The position of
SN 2005cs is marked as a white circle in the left panel.
The right panel shows the pre-SN NICMOS F110W image, with
the position of SN 2005cs marked with a $0\farcs09$ radius
circle, which represents the 3$\sigma$ uncertainty in the image registration.
We show a 2$\arcsec\times$2$\arcsec$ close-up of
the SN 2005cs progenitor environment on the pre-SN NICMOS
images in Figure 5. For comparison, the ACS F814W image
is shown in the left panel, with the progenitor of SN 2005cs
marked with an illustrative circle. The pre-SN NICMOS F110W and
F160W images are shown in the middle and right panels, with
a $0\farcs09$ radius circle that represents the 3$\sigma$ uncertainty in
the image registration.
Just outside the 3$\sigma$ error circle is a relatively
bright source in the NICMOS images, which we tentatively identify as
the counterpart of the star immediately to the northwest of the SN
progenitor seen in the F814W image. Note that the bright object immediately to
the southwest of the progenitor seen in the F814W image,
which we suggest is a blue compact cluster, has only a faint or undetected
counterpart in the near-infrared bands.


We marginally detect a source near the SN progenitor position in
the F110W image and, somewhat more suggestively,
in the F160W image.
However, this source seems to be offset slightly from the exact progenitor
position, suggesting it may be a different, very red star. However, the
NICMOS source is within 2--3$\sigma$ of the progenitor position, and
it is possible this source is the same as the star detected in the ACS
F814W image. To be conservative, we consider the brightness of this source
in the NICMOS images to be an upper limit to the flux of the SN
progenitor.

\section{Photometry of the Progenitor}

We have almost certainly identified the progenitor for
SN 2005cs in deep pre-SN {\it HST\/} images (see \S 4).
We attempt now to measure the brightnesses of this star
in the various bands.  From Figure 3 it is
apparent that SN 2005cs occurred in a crowded field; in particular,
the bright, blue object (likely a compact star cluster) immediately to
the southwest of the fainter star will complicate these measurements
in the ACS bands.

\subsection{ACS photometry}

We perform photometry on the ACS images following the prescription
outlined by Sirianni et al.~(2005).  Using the
IRAF/DAOPHOT package (Stetson 1987), we
combined the four individually ``drizzled'' images produced by the
{\it HST\/} ACS calibration pipeline in each of the four observed
passbands. We have confirmed that the photometric scale is consistent
between our combined images and the mosaic image produced by the
Hubble Heritage team. 

For the combined F814W and F555W images, we 
manually selected 377 and 384 (respectively) bright,
isolated stars to construct a spatially varying model
point-spread function (PSF) across the field.  Since photometric
accuracy for the progenitor would be improved if we could cleanly
remove the contribution of the neighboring bright compact cluster, we
carefully examined the characteristics of the cluster (i.e., its light
profile) and identified about 20--30 similar objects in each
image. Using our model stellar PSF, we first removed all the stars in
the neighborhood of our selected clusters.  Then, we constructed a
spatially constant model cluster ``point''-spread function (there are not
enough clusters to produce a spatially varying model), which was then fit to
the bright contaminating cluster.  Inspection of images in
which the cluster was subtracted revealed that the model fit did a
good job in removing its contribution in the F814W image, but
some small subtraction residuals remained in the F555W band.

Nonetheless, with the effect of the cluster minimized as much as
possible, we determined the brightness of objects in the SN environment
(within a radius of $0{\farcs}5$) using a $0{\farcs}15$-radius
aperture PSF (to maximize the signal-to-noise ratio)
fit in an iterative process, in which the brightest 2--3 stars were measured
first and subtracted away, then the next brightest stars, and so on.
By performing the photometry iteratively, we avoided potential errors
which could result from fitting all the bright and faint stars
simultaneously.  The progenitor's brightness was measured in the F814W
image when all other sources had been cleanly removed.

For the F555W image, since the progenitor is not detected, we
derived an upper limit to its brightness. We subtracted stars
of various magnitudes near the location of the progenitor (determined
from the F814W image) and visually inspected the subtracted images.
While this process involves some subjectivity and thus may have
large uncertainties, we nevertheless find that when stars brighter than
25.8 mag are subtracted from the F555W image, an apparent residual is
left in the subtracted images. We consider this magnitude (25.8) the
detection limit in the SN environment.  This limit
is brighter than the global limiting magnitude, $\age 26.5$ mag
(Mutchler \etal 2005), due to the presence of the neighboring cluster
and its imperfect subtraction.

We determined photometric corrections from the $0{\farcs}15$ PSF fitting
radius to a standard $0{\farcs}5$-radius aperture, using several isolated, bright
stars.  These are $0.19 \pm 0.02$ mag in F814W and $0.14 \pm 0.02$ mag in
F555W.  We then employed the tabulated correction to
infinite aperture of $0.092 \pm 0.001$ mag for the F555W image (Sirianni
\etal 2005).  Correcting the F814W photometry to infinite aperture was 
more involved: because both the PSF and aperture correction
at F814W depends on a star's
color, we first estimated an effective wavelength $\sim$8200~\AA\ for
the peak of the star's spectral energy distribution, based
on the spectral type suggested by the limit on the F555W$-$F814W
color for the progenitor. Following Sirianni \etal (2005), we derived 
a value for the correction, based on this wavelength; however, it
is quite similar to the tabulated
synthetic correction in Sirianni \etal (2005) of $0.087 \pm 0.001$ mag, which we
adopt for all the stars in the region.  Because of the fortuitous location of
the SN pixel position near a readout amplifier for the chip, the ACS charge
transfer efficiency correction is negligible for that position.

We then corrected the photometry for interstellar extinction. The
Galactic reddening along the line of sight to M51 is only $E(B-V)$ =
0.035 mag (Schlegel \etal 1998); however, the presence of Na~I~D line absorption
in the SN spectrum at the redshift of M51 suggests some additional
host-galaxy extinction toward the SN (Modjaz 2005). We therefore adopted a total
reddening $E(B-V)$ = 0.10$\pm$0.05 mag (Richmond 2005) toward the SN progenitor.
For the other stars in the SN environment, we lack constraints on the
host extinction and corrected only for the Galactic component.  Finally,
we transformed the F814W and F555W magnitudes to Johnson-Cousins $V$ and $I$
(Sirianni \etal 2005) and obtained $I = 24.15 \pm 0.20$, $V > 25.5$ mag for
the progenitor.  Adopting a distance modulus $\mu = 29.6 \pm 0.3$
mag (Richmond \etal 1996), the progenitor has $M_I^0 \approx -5.5$ mag and
$(V-I)^0 \age 1.3$ mag, consistent with a red supergiant of spectral
type K0 or later.

Since the F435W and F658N images are not of sufficiently high
signal-to-noise ratio to derive meaningful limits on the progenitor
brightness (the progenitor is not
detected in either image) and enhance our knowledge of the progenitor
beyond what we have now learned from the F555W and F814W images,
we do not consider them further.

\subsection{NICMOS Photometry}

We performed photometry on the subsampled drizzled
NICMOS F110W and F160W pre-SN images, with resolution
$0\farcs1$ pixel$^{-1}$. The photometric zero points appropriate for
Cycle 7, when these NICMOS observations
were obtained, were adopted. Since the SN progenitor
is, at best, marginally detected, and it occurred in
a crowded region, we employed a procedure for the photometry
similar to that described above, in \S 3.1, for the ACS F555W image:
stars were subtracted iteratively in order of decreasing brightness
near the SN location, using the appropriate
TinyTim model PSFs (Krist \& Hook 2003)
for the NICMOS filters, until the visually most satisfactory
subtracted image was achieved.  As a result, we derived
$J$ = 22.5$\pm$0.5 and $H$ = 21.5$\pm$0.5 mag for the marginally
detected emission close to the SN site, and adopted these magnitudes for
the limiting brightness of the progenitor in each band.

\section{Discussion}

\subsection{Is SN 2005cs of Type II-Plateau?}

Type II SNe can be further divided into several subclasses;
see the review of SN types in Filippenko (1997) for more details and
references. The two classical subtypes are the Type II-plateau
(SNe II-P), with a pronounced plateau phase seen in their optical light
curves, and the Type II-linear (SNe II-L),
with a linear decline after their maximum brightness.
Additional subtypes include the
Type II-narrow (SNe IIn),
with narrow emission lines in their spectra (often, but not
always, superimposed on a broader emission component), and the
SN 1987A-like (peculiar, subluminous, with a unique photometric
behavior).  Moreover, the Type IIb SNe, such as SN 1993J in M81
(Filippenko \etal 1993; Nomoto \etal 1993),
manifest themselves as SNe~II at early times, but then experience a
metamorphosis into a Type Ib SN at late times.  To some extent the
shapes of the optical light curves correlate with the spectral subtype
for SNe II.

%

We have been able to follow SN 2005cs in \ubvri\
with KAIT roughly every other night since its discovery, sampling
the light curves for $\age$20~d since explosion, so
we should already be getting an indication of the SN subtype.
Unfortunately, the absolute calibration for SN 2005cs has not yet been
established, but in Figure 5 we show its \ubvri\ light curves 
relative to the recent prototypical SNe II-P 1999em (Leonard \etal 2002b;
Hamuy \etal 2001) and 2004et (Li \etal 2005a).
The SN 2005cs light curves were obtained by performing differential
aperture photometry between the SN and two or more bright stars
in the field.  In the figure we show the curves adjusted in magnitude
to match the light curves for SN 1999em and SN 2004et; the number of 
days since explosion has not
been adjusted. The final photometry for SN 2005cs
will be published in a forthcoming paper, when proper absolute calibration
for the field has been obtained, along with analysis of available optical
spectra.

Even from these relative light curves, it can
be clearly seen that SN 2005cs is currently undergoing a plateau phase,
especially in the $V$, $R$, and $I$ bands.  SN 2005cs also follows very
similar photometric evolution to that of SN 1999em and SN 2004et in these
three bands, but seems to be evolving somewhat
faster in the $U$ and $B$ bands. At the distance of M51,
SN 2005cs had a peak $V$-band magnitude on the plateau of only $-15.6$,
so it is rather subluminous. SN 2005cs also had a relatively
small early expansion velocity,  $\sim$7500 km s$^{-1}$, derived from
the absorption minimum of H$\beta$ (Modjaz \etal 2005).
We believe there should be little doubt
that SN 2005cs is a SN II-P, albeit possibly somewhat unusual and subluminous,
relative to the prototypes.

\subsection{An X-ray Flash from SN 2005cs?}

An X-ray flash was detected in a 2136~s {\it Swift}/XRT observation
of M51 on 2005 July 6.231, 8 days after the discovery of SN 2005cs
(Immler \etal 2005). This generated interest as to whether or not the
flash was related to SN 2005cs. The X-ray flash is a 6$\sigma$
detection, and is located about 10$\arcsec$ from the SN 2005cs
position. No X-ray source was detected 96 minutes later
in a 2031~s exposure, as well as in all other 
{\it Swift}/XRT observations of M51 on June 30, July 3, 5, 6
and 7. Given the timing of the X-ray flash
($>$8 days after the SN explosion), and the relative offset of the detection
from the SN position, speculation emerged that the new source was
unrelated to SN 2005cs. However, since the XRT PSF has a half-power
diameter of 18$\arcsec$ at 1.5~keV, based on position alone the possibility
that the emission is from SN 2005cs cannot be excluded.

From the photometry obtained with the Palomar Observatory robotic
60-in telescope, Gal-Yam (2005) reports that no photometric
anomaly was detected for SN 2005cs around the time of the
detection of the X-ray flash in M51, and yet no additional variable
sources were detected within 20$\arcsec$ of SN 2005cs.
We obtained with KAIT a sequence of \ubvri\ images for
SN 2005cs, starting on July
6.246, just 20 minutes after the {\it Swift}/XRT detection of the X-ray
flash. The photometry from and timing of these images are indicated
with an arrow in Figure 5. Our data are consistent with what Gal-Yam (2005)
found: the SN evolved photometrically as expected, following the trend
before and after the X-ray flash detection, particularly in $U$.

We further compare the ACS/HRC images of the SN 2005cs field in the UV
bands taken on July 11.501, 5.27 days after the X-ray flash detection,
to the pre-SN ACS/WFC F435W ($\sim B$) images. We do not detect any apparent
new sources within 10$\arcsec$ of SN 2005cs, to limiting magnitudes
$\sim$22.5.

Thus, either the X-ray flash was unrelated to SN 2005cs, or it did
not produce any detectable anomalous behavior at optical wavelengths 
shortly thereafter. The nature of the X-ray flash remains a mystery.

\subsection{The Spectral Type and Mass of the Progenitor}

We can further constrain the spectral type of the progenitor of SN II-P 
2005cs, a red supergiant, from the {\sl HST\/} photometry.  
In Figure 6 we show the star's implied spectral
energy distribution (SED); obviously, the star was detected only
in $I$, so we can place only upper limits on its $VJH$ brightness in this
diagram.
The $I$ brightness and the upper limits have been corrected for reddening.
Also shown for comparison are SEDs of some late-type
supergiants (Drilling \& Landolt 2003; Tokunaga 2003), reddened by the
assumed $E(B-V)$ to the SN and all normalized to the progenitor's $I$ magnitude.
We see that the $V$ upper limit constrains the spectral type to later than about
K0, while the $JH$ upper limits constrain it to earlier than M5.

We can further venture to estimate the initial mass of this
star by comparing its intrinsic brightness and color limit
to theoretical massive stellar
evolutionary tracks.  Lejeune \& Schaerer (2001) have generated tracks
for a range of zero-age main sequence (ZAMS) masses for several
different metallicities and assuming enhanced mass loss
for the most massive stars.
From metallicity measurements in
M51 (Zaritsky \etal 1990), we estimate that at the SN 2005cs site,
$12 + \log$ [O/H] = 9.06 $\pm$ 0.04 dex. This is 0.26 dex higher
than the solar abundance (Grevesse \& Sauval 1998), implying that the
metallicity in the SN environment could be somewhat higher than solar.
We therefore consider the tracks from Lejeune \& Schaerer for
both $Z = 0.02$ (i.e., solar) and $Z = 0.04$ (the next higher metallicity).


In Figure 7 we show the intrinsic color-magnitude diagram for the SN
environment (the progenitor and other stars within a $0{\farcs}5$ radius
of the SN) and overlay the tracks for the two possible metallicities
for a variety of stellar masses.
The total photometric uncertainties shown are the
measurement uncertainties and the uncertainty in the distance
modulus added in quadrature.  It is also clear from the diagram
that the SN environment is abundant in red supergiants with masses of
7--15~$M_\odot$, although from the astrometric arguments above, we
have eliminated these stars as potential candidates for the
progenitor.  

The location in the diagram of the progenitor itself suggests
that it had $M_{\rm ZAMS} \approx 7$--9~$M_\odot$.  This estimated
mass is right at the $\sim 8 M_\odot$ theoretical
lower limit for core collapse in massive stars (Woosley \& Weaver 1986).
Furthermore, we consider it highly unlikely that we have {\it not\/}
correctly identified the SN progenitor, if it is indeed a core-collapse
SN. The low reddening to the SN and the
limiting magnitude in the F814W image, $M_I^0 \approx -4$ mag, 
imply that any star which has escaped detection, yet somehow
remains a possible progenitor candidate, has
a ZAMS mass ($< 7 M_\odot$) formally below the core-collapse limit.

\subsection{Implications from the SN 2005cs Progenitor}

We consider our identification of the progenitor of SN 2005cs quite 
secure and the estimates for its inferred spectral type and ZAMS mass 
compelling.
SN 2005cs is only the third SN II-P to have its progenitor directly
identified on pre-SN images.  The other two are SN 2003gd (Van Dyk \etal 2003;
Smartt \etal 2004; Hendry \etal 2005), which also had a
progenitor with initial mass (7--9 $M_\odot$) very near the theoretical
limit for core collapse, and SN 2004et, with an initial mass in the range
13--20 $M_\odot$ (Li \etal 2005a).  The former was a subluminous SN,
also with a red supergiant progenitor, and the latter may have been unusual,
with apparently a yellow supergiant progenitor.  

Upper limits on the initial masses of other, well-studied SNe~II-P based 
on pre-SN images have also been established:
$M_{\rm ZAMS} \leq 15 M_\odot$ for SN 1999em (Smartt \etal 2002),
$\leq 15 M_\odot$ for SN 1999gi (Leonard et al. 2002a), $\leq 13 M_\odot$
for SN 2001du (Van Dyk \etal 2003b; Smartt \etal 2003),
and $\leq 12 M_\odot$ for SN 1999br (Maund \& Smartt 2005).  From these
detections and upper limits, a trend is emerging for SNe II-P, the most
common core-collapse SNe, that the majority (if not all) of them
arise from stars with masses in the range $\sim$8--15 $M_\odot$.
We note that a progenitor in the mass range 20--40 $M_\odot$ has yet to be
found for a normal SN~II-P, and thus the fate of these very massive stars
still needs to be observationally verified. However, because of the
steeply declining mass function, it may be quite some time before a dearth
of very massive progenitors presents a significant challenge to theory.

It is also unclear whether stars more massive than 20~$M_\odot$
actually become normal SNe~II-P, or give rise to SNe~II-L, such as 
SN 1979C (Branch \etal 1981; Van Dyk \etal 1999a), or SNe~IIn, such 
as SN 1988Z (Stathakis \& Sadler 1991; Chugai \& Danziger 1994)
and SN 1995N (Fransson \etal 2002). Massive blue variable stars
($\age 30$--40 $M_\odot$) may undergo super-outbursts (sometimes 
misclassified as SNe) while in the luminous blue variable stage
(e.g., Van Dyk \etal 2000), en route to becoming Wolf-Rayet stars, 
SNe~Ic, and perhaps even collapsars (which appear to be responsible for 
long-duration gamma-ray bursts).

We have already discussed that SN 2005cs could be subluminous, as
was SN 2003gd (Van Dyk \etal 2003; Hendry \etal 2005).
Its post-plateau photometric behavior will corroborate this (we predict
it should decline markedly in brightness in all bands after the plateau phase,
much more so than for prototypical SNe II-P).
Zampieri \etal (2003) proposed that these subluminous SNe~II-P with
low expansion velocities and low $^{56}$Ni yields, such as SN 1999br, 
SN 2003gd, and now possibly SN 2005cs, originate from high-mass ($\geq
20 M_\odot$) progenitors in which the rate of early infall of stellar material
on the collapsed core is large. They further postulated that events
of this type could form a black hole remnant, giving rise to significant
fallback and late-time accretion. It is apparent from the direct
mass estimates for both the SN 2003gd and SN 2005cs progenitors, as well
as from the upper mass limit for SN 1999br, that these subluminous SNe 
are produced by relatively low-mass red supergiants, 
which may end their lives in less energetic explosions.

We also note that the low energy output and the low synthesized yields
for $^{56}$Ni  suggest that, even though the rate of such
events is higher than SNe from more massive stars, the impact on
the energy input and chemical evolution of their host galaxies per event is
relatively small.  Certainly, it is thought that at very early times in galactic
history ($z \age 6$), the initial mass function was heavily skewed to
very massive stars of essentially zero metallicity, and these stars contributed
entirely to the early galactic evolution (e.g., Matteucci \& Calura 2005;
Umeda \& Nomoto 2005).
The contribution of lower-mass stars
has increased steadily over cosmic history, although the explosions of the
more massive stars could still have a large impact on galaxies today.
A careful SN rate
calculation such as that being conducted by Leaman \etal (2004),
and detailed models for nucleosynthesis in the SN~II explosions
(Woosley \& Weaver 1995), will provide useful information
on the relative contribution to the chemical evolution of galaxies
from SNe of various progenitor masses.

\section{Conclusions}

In this paper we have analyzed {\it HST}/ACS and NICMOS
data for M51 before the discovery of SN 2005cs,
and we have identified the progenitor of SN 2005cs. The secure identification
of the progenitor is achieved by geometrical transformations, using
new images of SN 2005cs from KAIT, CFHT, and {\it HST}/ACS.
We measure $I$ = 24.15$\pm$0.20, $V > 25.5$, $J> 22.5$, and
$H>21.5$ mag for the SN progenitor.
Together, this information suggests that
SN 2005cs originated from a red supergiant of spectral type
K0--M3, with an initial mass of 7--9~$M_\odot$.  A significant trend appears
to be emerging, that SNe~II-P arise from massive stars with initial
masses $\sim 8$--15~$M_{\odot}$.

SN 2005cs is a very important addition to those SNe whose progenitors
have been directly identified, a very small (but growing)
number of objects.  This identification would have been very
difficult without the deep images of superior spatial resolution obtained
with {\it HST}/ACS; the progenitor star would almost certainly have been lost
in the light of the neighboring bright star cluster had only inferior images
been available.  We will continue to monitor SN 2005cs with KAIT
and other telescopes to better define the nature of the SN itself; it remains
to be seen how this SN will ultimately behave.  



\acknowledgments

The work of A.V.F.'s group at U.C. Berkeley is supported by National
Science Foundation grant AST-0307894, and by NASA/{\it HST} grant
GO-10182 from the Space Telescope Science Institute, which is operated
by the Association of Universities for Research in Astronomy, Inc.,
under NASA contract NAS 5-26555. He is also grateful for a Miller
Research Professorship at U.C.B., during which part of this work was
completed. We thank Stefan Immler for his private communications on 
the X-ray flash in M51. KAIT was made possible by generous donations from Sun
Microsystems, Inc., the Hewlett-Packard Company, AutoScope
Corporation, Lick Observatory, the NSF, the University of California,
and the Sylvia \& Jim Katzman Foundation.

\clearpage

\newpage

\begin{figure*}
\centerline{\psfig{file=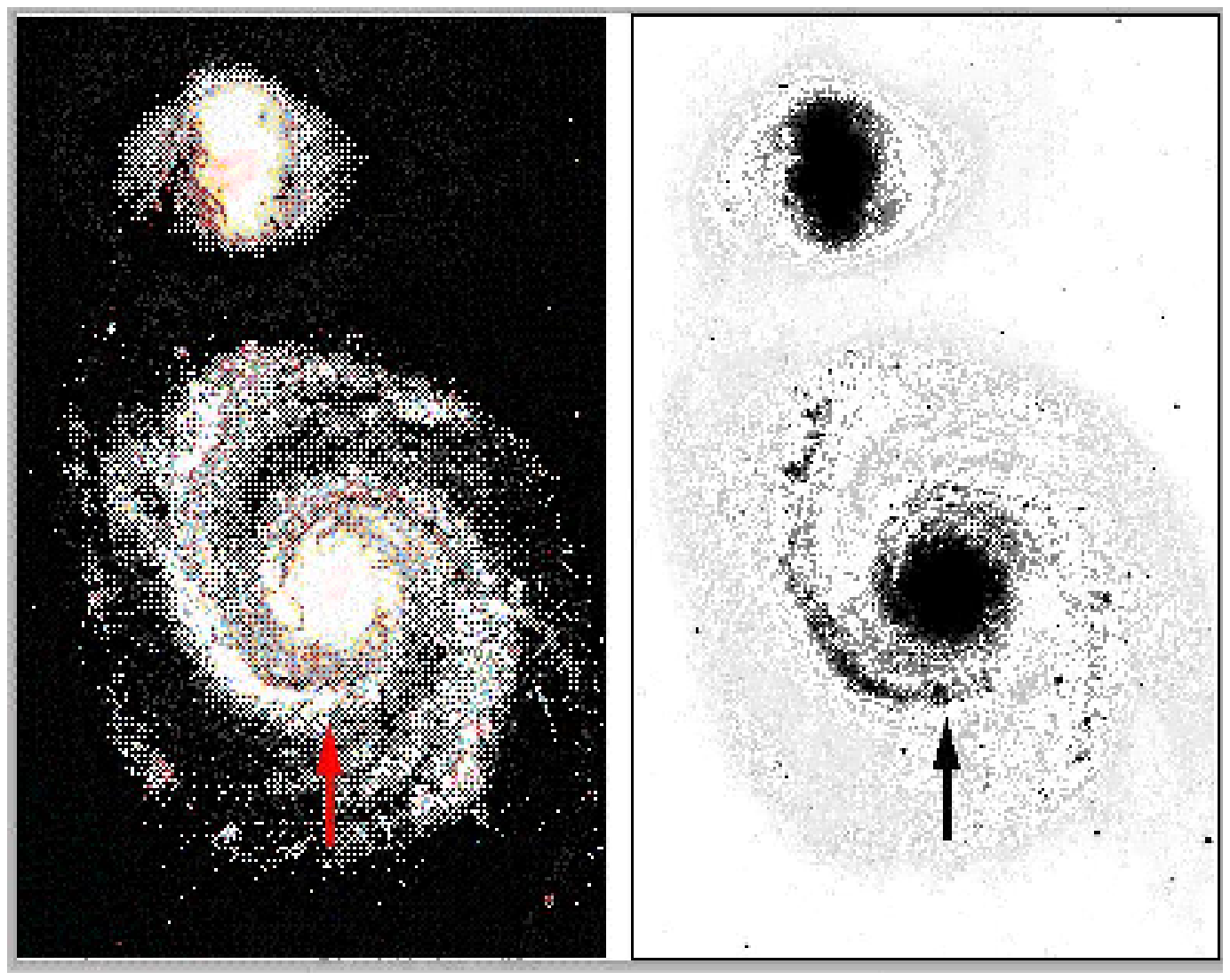,width=7.0in,angle=270}}
\caption[] {

Comparison of the Whirlpool Galaxy (M51) before and after the SN 2005cs
explosion. North is up, and east is to the left. The total field of
view for each image is $5{\farcm}96 \times 9{\farcm}56$. {\bf Left:}
The color composite {\it HST\/} ACS mosaic image. The position of the
SN 2005cs progenitor is marked by a red arrow.  {\bf Right:} The same
field as imaged by the MegaCam instrument on the 3.6-m
Canada-France-Hawaii Telescope (CFHT) on 2005 Jul 2.28.
SN 2005cs is marked by a black arrow.

}
\end{figure*}

\newpage

\begin{figure*}
\centerline{\psfig{file=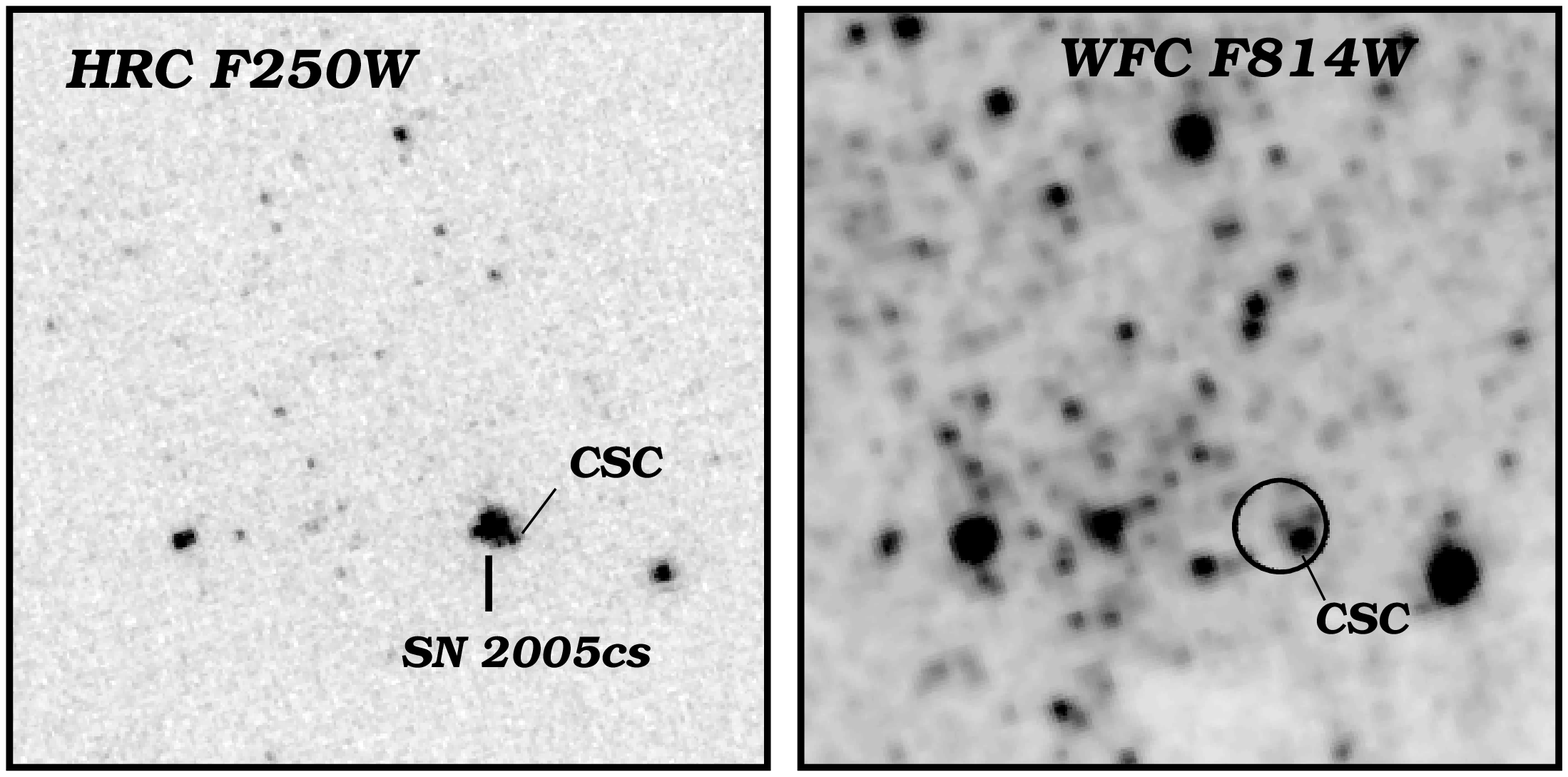,width=6.3in,angle=0}}
\caption[] {

Comparison of a $5\arcsec \times 5\arcsec$ region in the ACS
images before and after SN 2005cs. North is up, and east is to
the left. {\bf \it Left}: A 480-s F250W image obtained with the HRC on 2005
July 11.50. The position of SN 2005cs is marked. {\bf \it Right}:
The F814W image obtained with the WFC five months prior to the SN 2005cs explosion.
An illustrative $0\farcs3$ radius circle is centered on
the location of the SN position: the true geometrical transformation
error is only $0\farcs005$. The object located at the center
of the circle is the progenitor of SN 2005cs. The object labeled
``CSC'' to the southwest of the progenitor is probably a 
compact star cluster.}

\label{fig:closeup}
\end{figure*}

\newpage

\begin{figure*}
\centerline{\psfig{file=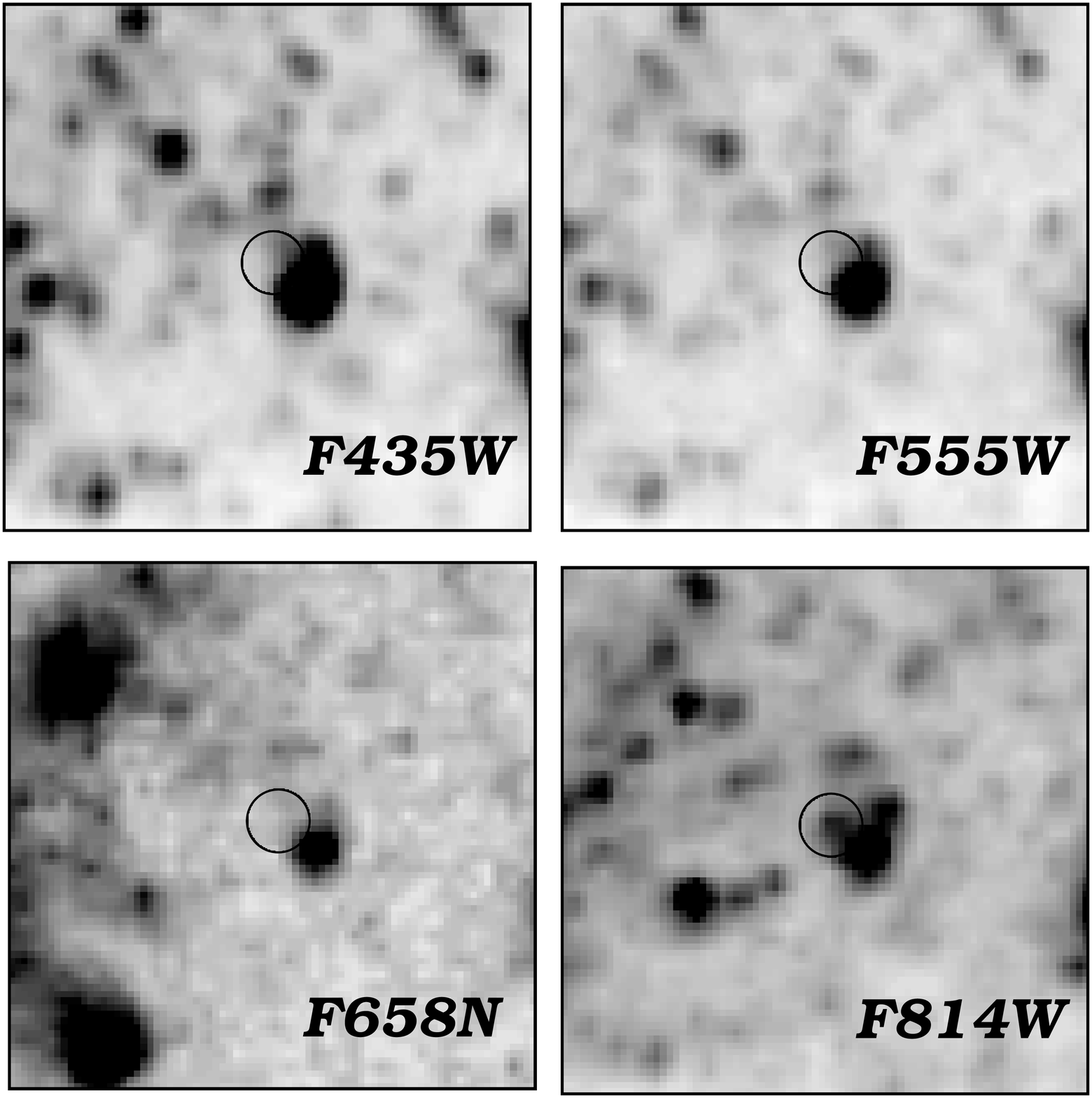,width=6.3in,angle=0}}
\vskip -7cm
\caption[] {

A $2\arcsec \times 2\arcsec$ close-up of the SN 2005cs environment in
the {\it HST}/ACS F435W, F555W, F658N, and F814W images.  North is up,
and east is to the left in each panel.  The position of the progenitor
measured from the F814W image is marked as a circle in each panel.
The progenitor is apparently not detected in the F435W, F555W,
or F658N images. A bright, blue object immediately to
the southwest of the SN 2005cs progenitor is likely to be a compact
star cluster. }
\label{fig:closeup}
\end{figure*}

\newpage

\begin{figure*}
\centerline{\psfig{file=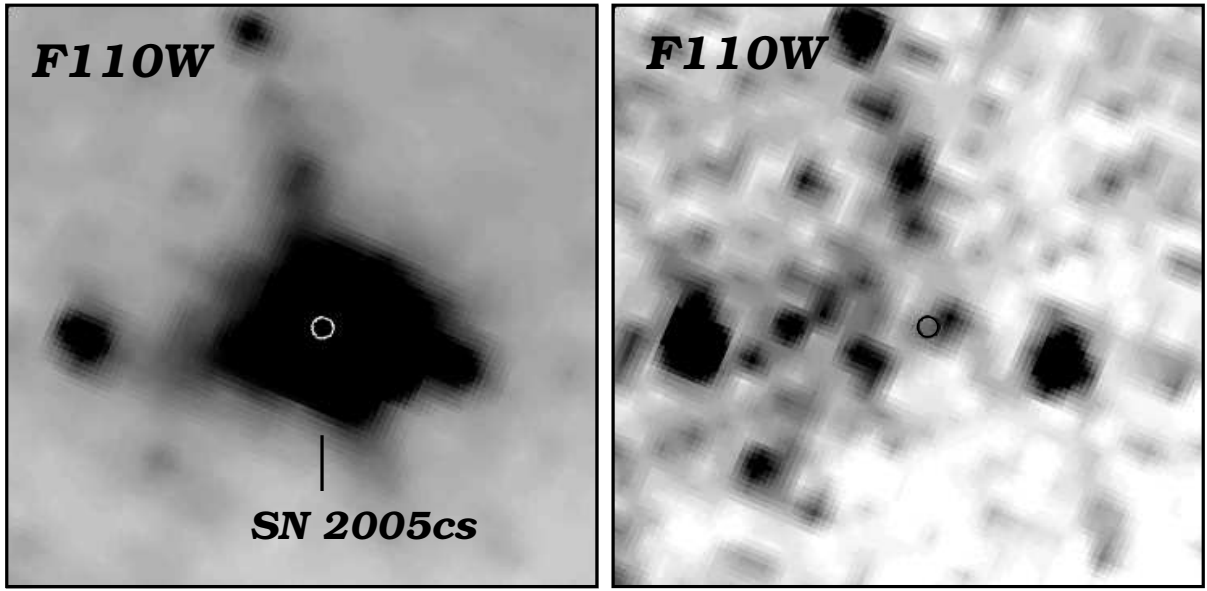,width=6.3in,angle=0}}
\caption[] {
Comparison of a $5\arcsec \times 5\arcsec$ region in the NICMOS
images after and before SN 2005cs. North is up, and east is to
the left. {\bf \it Left}: A 31.97-s F110W image obtained with the
NICMOS/NIC3 on 2005
July 13.82. The position of SN 2005cs is marked with a white
circle. {\bf \it Right}:
A 31.97-s F110W image obtained with the NICMOS/NIC3 prior to the SN 2005cs
explosion.
A $0\farcs09$-radius circle is centered on
the location of the SN position, which represents the
3$\sigma$ uncertainty in the geometrical transformation.

}
\label{fig:closeup}
\end{figure*}

\newpage

\newpage

\begin{figure*}
\centerline{\psfig{file=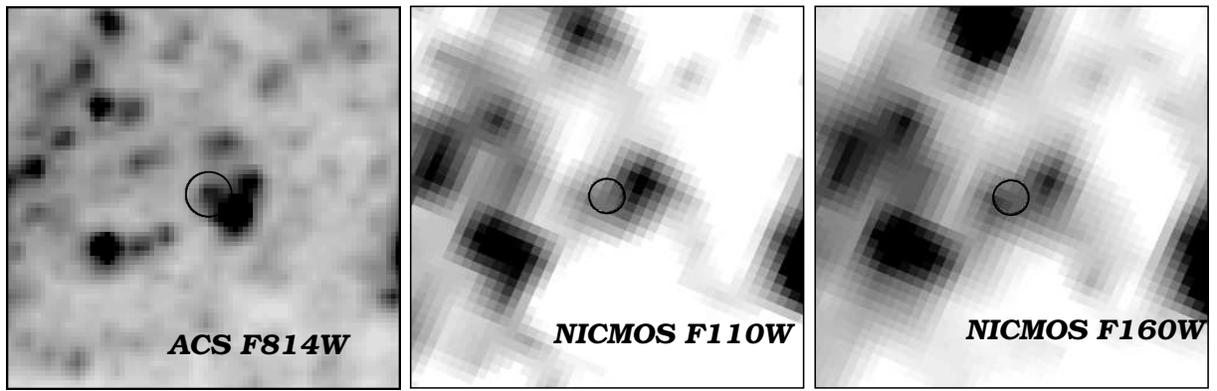,width=6.3in,angle=0}}
\caption[] {
The {\it HST}/NICMOS images of the SN 2005cs site. Each panel is
$2\arcsec \times 2\arcsec$. North is up, and east is to the left.
{\bf \it Left:} The ACS/F814W image, with the position of the progenitor
marked by an illustrative circle. {\bf \it Middle:} The pre-SN NICMOS/F110W
image. The position of the SN 2005cs progenitor is marked with a circle
that represents a 3$\sigma$ positional uncertainty when the pre-SN
NICMOS/F110W image is registered to the new NICMOS/F110W image showing
SN 2005cs (see Figure 4 and the text for details). {\bf \it Right:} The same as the
middle panel, but for the NICMOS/F160W image.

}
\label{fig:cmd}
\end{figure*}

\newpage

\begin{figure*}
\centerline{\psfig{file=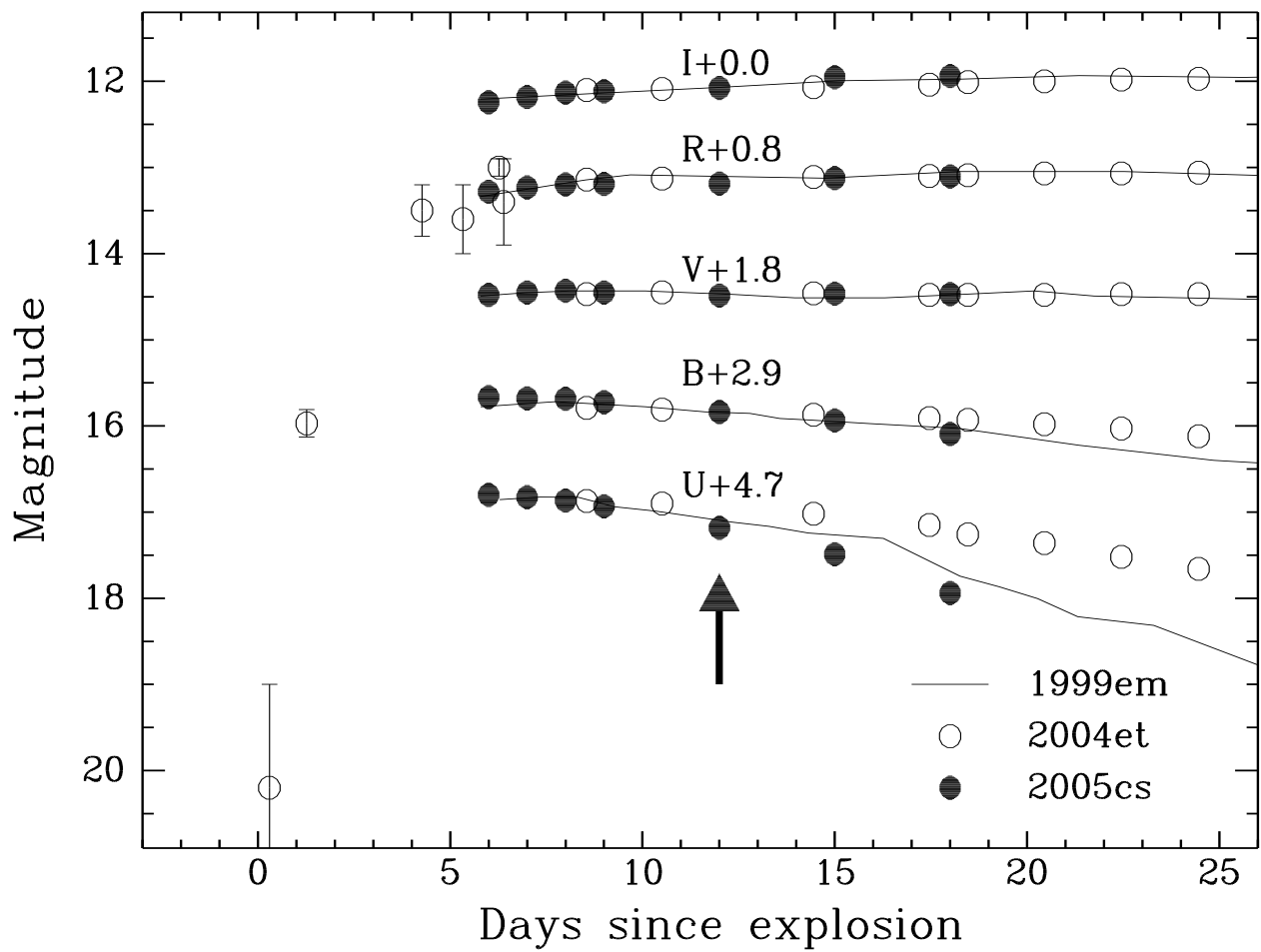,width=7.3in,angle=270}}
\caption[] {
Preliminary KAIT light curves for SN 2005cs ({\it solid circles}).
Also overplotted are light curves of SN 1999em ({\it solid line}; Leonard
\etal 2002b; Hamuy \etal 2001) and
SN 2004et ({\it open circles}; Li \etal 2005). The magnitudes shown for SN 2005cs
are not on an absolute scale (see text for details). The light curves for SNe
have all been shifted in magnitude (but not in date since explosion) to match
each other. Also marked with an arrow is the time when the X-ray flash was detected in
M51 (Immler \etal 2005).

}
\label{fig:lc}
\end{figure*}

\newpage

\begin{figure*}
\centerline{\psfig{file=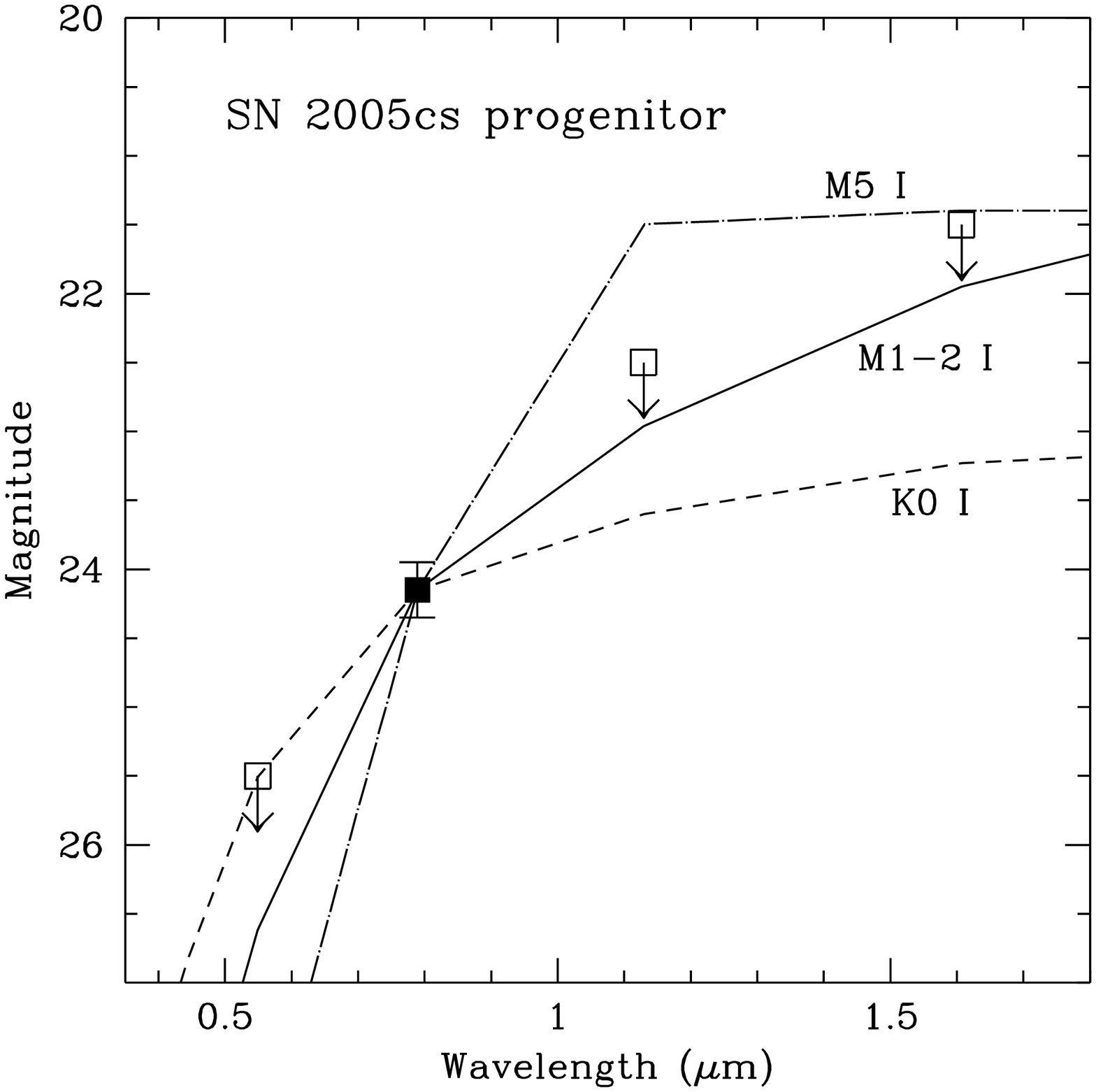,width=6.3in,angle=0}}
\caption[] {

Spectral energy distribution (SED) inferred for the SN 2005cs progenitor,
based on photometry of the pre-SN {\sl HST\/} ACS and NICMOS images
(see text).  For comparison we show the SEDs for K0, M1--2, and M5
supergiants (Drilling \& Landolt 2000; Tokunaga 2000), all reddened by
$E(B-V)=0.1$ mag and all adjusted to
$I=24.15$ mag.  The photometric upper limits derived from these
{\sl HST\/} images constrain the progenitor's spectral type
to approximately between K0 and M3.

}

\label{fig:cmd}
\end{figure*}

\newpage

\begin{figure*}
\centerline{\psfig{file=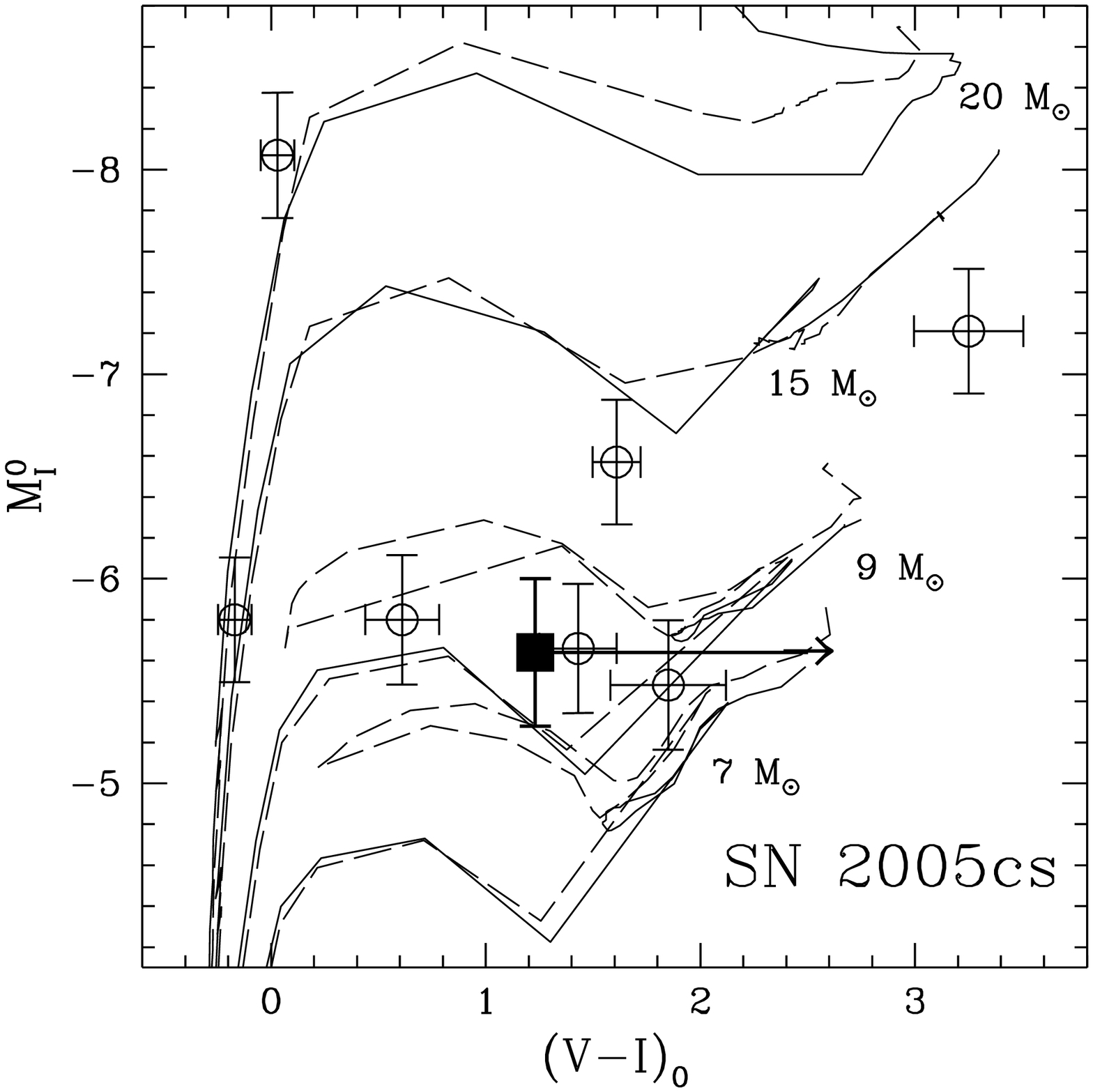,width=6.3in,angle=0}}
\caption[] {

The $(V-I)^0$ vs. $M^0_I$ color-magnitude diagram of the SN 2005cs
environment. The progenitor of SN 2005cs is shown as a {\it filled
square}, while the other objects in the neighborhood of the progenitor
are plotted as {\it open circles}.  For comparison we show
model stellar evolutionary tracks for a range of masses (Lejeune \&
Schaerer 2001),
assuming enhanced mass loss for the most massive stars, with solar
metallicity ($Z = 0.02$, {\it dashed line}) and above-solar
metallicity ($Z = 0.04$, {\it solid line}).  The location of SN
2005cs in this diagram suggests that it was a red supergiant with ZAMS
mass 7--9 $M_\odot$.  }
\label{fig:cmd}
\end{figure*}

\newpage

\begin{deluxetable}{ccccccr}
\tablecaption{{\it HST}/ACS and NICMOS Data for M51 Pre-SN 2005cs}
\tablehead{
\colhead{Dataset} & \colhead{UT Date} &\colhead{Exp. (s)}
&\colhead{Inst.} &\colhead{Aperture} &\colhead{Filter}
&\colhead{Prop. ID}
}
\startdata
h\_m51\_b\_s05\_drz\_sci\tablenotemark{a}  & 2005 Jan & 2720 & ACS & WFC & F435W & 10452 \\
h\_m51\_v\_s05\_drz\_sci\tablenotemark{a}  & 2005 Jan & 1360 & ACS & WFC & F555W & 10452 \\
h\_m51\_h\_s05\_drz\_sci\tablenotemark{a}  & 2005 Jan & 1360 & ACS & WFC & F658N & 10452 \\
h\_m51\_i\_s05\_drz\_sci\tablenotemark{a}  & 2005 Jan & 1360 & ACS & WFC & F814W & 10452 \\
N48R02150 &1998 Jun 28 & 255.72&NICMOS & NIC3&F110W&7237\\
N48R02160 &1998 Jun 28 & 255.72&NICMOS & NIC3&F160W&7237\\
N48R02170 &1998 Jun 28 & 255.72&NICMOS & NIC3&F222M&7237\\
N48R02180 &1998 Jun 28 & 511.61&NICMOS & NIC3&F187N&7237\\
N48R02190 &1998 Jun 28 & 511.61&NICMOS & NIC3&F190N&7237\\
\enddata
\tablenotetext{a}{This image is the final mosaic of many single exposures.}
\end{deluxetable}

\begin{deluxetable}{ccccccr}
\tablecaption{{\it HST}/ACS and NICMOS Observations of SN 2005cs}
\tablehead{
\colhead{Dataset} & \colhead{UT Date} &\colhead{Exp. (s)}
&\colhead{Inst.} &\colhead{Aperture} &\colhead{Filter}
&\colhead{Prop. ID}
}
\startdata
J90ZP1010 & 2005 Jul 11& 960 & ACS & HRC & F220W & 10182 \\
J90ZP1020 & 2005 Jul 11& 480 & ACS & HRC & F250W & 10182 \\
J90ZP1030 & 2005 Jul 11& 240 & ACS & HRC & F330W & 10182 \\
N90ZQ1020 & 2005 Jul 13& 255.72& NICMOS & NIC3& F110W & 10182 \\
N90ZQ1030 & 2005 Jul 13& 511.71& NICMOS & NIC3& F160W & 10182 \\
\enddata
\end{deluxetable}


\newpage

\begin{thebibliography}{}
\bibitem[Aldering et al.~1994]{ald94} Aldering, G., Humphreys, R.~M., \&
Richmond, M. W. 1994, \aj, 107, 662
\bibitem{}Branch, D., Falk, S.~W., Uomoto, A.~K., Wills, B.~J., McCall,
M.~L., \& Rybski, P. 1981, \apj, 244, 780
\bibitem{}Chugai, N. N., \& Danziger, I. J., 1994, \mnras, 268, 173
\bibitem[Drilling \& Landolt 2000]{dri00} Drilling, J. S., \& Landolt, A. U.
2000, in Allen's Astrophysical Quantities, 4th ed., ed.~A. N. Cox (New York:
AIP), 381
\bibitem{}Filippenko, A. V. 1997, \araa, 35, 309
\bibitem[Filippenko et al.~2001]{fil01}Filippenko, A. V., Li, W., Treffers,
R. R., \& Modjaz, M. 2001, in Small-Telescope Astronomy on Global Scales,
ed. W.-P. Chen, C. Lemme, \& B. Paczy\'{n}ski (ASP Conf. Ser. 246; San
Francisco: ASP), 121
\bibitem{}Filippenko, A. V., Matheson, T., \& Ho, L. C. 1993, \apj, 415, L103
\bibitem{}Fransson, C., \etal 2002, \apj, 572, 350
\bibitem{}Gal-Yam, A. 2005, ATEL, No. 561
\bibitem[Gilmozzi et al.~1987]{gil87} Gilmozzi, R., et al. 1987, Nature, 328,
318
\bibitem[Grevesse \& Sauval 1998]{gre98} Grevesse, N., \& Sauval, A. J. 1998,
Space Sci. Rev., 85, 161
\bibitem[Hamuy et al.~2001]{ham01} Hamuy, M., et al. 2001, \apj, 558, 615
\bibitem{}Hendry, M. A., \etal 2005, \mnras, 359, 906
\bibitem{}Immler, S., Kong, A., \& Lewin, W. H. G. 2005, IAU Circ., No. 8564
\bibitem{}Kloehr, W. 2005, IAU Circ., No. 8553
\bibitem{}Krist, J., \& Hook, R. 2003, Tiny Tim User's Manual, v6.1
\bibitem{}Leaman, J., Li, W., \& Filippenko, A. V. 2004, AAS Meeting 205, \#71.02
\bibitem[Lejeune \& Schaerer 2001]{lej01} Lejeune, T., \& Schaerer, D. 2001,
\aap, 366, 538
\bibitem{}Leonard, D.~C., et al., 2002a, \aj, 124, 2490
\bibitem[Leonard et al.~2002a]{leo02a} Leonard, D.~C., et al., 2002b, \pasp,
114, 35
\bibitem[Li et al.~2000]{wli00} Li, W., et al. 2000, in Cosmic Explosions,
ed. S.~S. Holt \& W.~W.~Zhang (New York: American Institute of Physics), 103
\bibitem{}Li, W. 2005, IAU Circ., No. 8553
\bibitem{}Li, W., Filippenko, A.~V., \& Van Dyk, S.~D. 2005b, ATEL, No. 492
\bibitem{}Li, W., Filippenko, A.~V., \& Van Dyk, S.~D.. 2005e, IAU Circ., No. 8565
\bibitem{}Li, W., Van Dyk, S.~D., \& Filippenko, A.~V. 2005c, IAU Circ., No. 8556
\bibitem{}Li, W., Van Dyk, S.~D., Filippenko, A.~V., \& Cuillandre, J.-C. 2005a,
\pasp, 117, 121
\bibitem{}Li, W., Van Dyk, S.~D., Filippenko, A.~V., Cuillandre, J.-C., \& Jha, S. 
2005d, IAU Circ., No. 8556
\bibitem{}Matteucci, F., \& Calura, F. 2005, \mnras, 360, 447
\bibitem{}Maund, J.~R., \& Smartt S.~J. 2005, \mnras, 360, 288
\bibitem{}Modjaz, M., Kirshner, R. P., Challis, P., \& Hutchins, R. 2005, IAU Circ., No. 8555
\bibitem{}Mutchler, M., \etal 2005, BAAS, Vol. 37, No. 2
\bibitem{}Richmond, M.~W. 2005, IAU Circ., No. 8555
\bibitem{}Richmond, M.~W., \etal 1996, \aj, 111, 327
\bibitem[Ryder et al.~1993]{ryd93} Ryder, S., Staveley-Smith, L., Dopita, M.,
Petre, R., Colbert, E., Malin, D., \& Schlegel, E. M. 1993, \apj, 416, 167
\bibitem[Schlegel, Finkbeiner, \& Davis 1998]{sch98} Schlegel, D. J.,
Finkbeiner, D. P., \& Davis, M. 1998, \apj, 500, 525
\bibitem{}Sirianni, M., \etal 2005, \pasp, submitted.
\bibitem{}Smartt, S. J., Gilmore, G. F., Tout, C. A., \& Hodgkin, S. T 2002,
\apj, 565, 1089
\bibitem[Smartt, Maund, \& Hendry 2003]{sma03b} Smartt, S. J.,
  Maund, J. R., Hendry, M. A., Tout, C. A., Gilmore, G. F., Mattila, S.,
  \& Benn, C. R. 2004, Science, 303, 499
\bibitem{}Smartt, S. J., \etal 2003, \mnras, 343, 735
\bibitem[Sonneborn, Altner, \& Kirshner 1987]{son87} Sonneborn, G., Altner, B.,
\& Kirshner, R.~P. 1987, \apjl, 323, L35
\bibitem{}Stathakis, R. A., \& Sadler, E. M. 1991, \mnras, 250, 786
\bibitem[Stetson 1987]{ste87} Stetson, P. B. 1987, \pasp, 99, 191
\bibitem[Tokunaga 2000]{tok00} Tokunaga, A.T. 2000, in Allen's Astrophysical
Quantities, 4th ed., ed.~A. N. Cox (New York: AIP), 143
\bibitem{}Umeda, H., \& Nomoto, K. 2005, \apj, 619, 427
\bibitem{}Van Dyk, S. D., Li, W., \& Filippenko, A. V. 2003a, \pasp, 115, 1289
\bibitem{}Van Dyk, S. D., Li, W., \& Filippenko, A. V. 2003b, \pasp, 115, 448
\bibitem{}Van Dyk, S. D., Garnavich, P. M., Filippenko, A. V., H\"{o}flich, P.,
   Kirshner, R. P., Kurucz, R. L., \& Challis, P. 2002, \pasp, 114, 1322
\bibitem[Van Dyk et al.~1999b]{van99b} Van Dyk, S. D., Peng, C. Y., Barth,
  A. J., \& Filippenko, A. V. 1999, AJ, 118, 2331
\bibitem[Van Dyk et al.~1999a]{van99a} Van Dyk, S. D., Peng, C. Y., Barth,
  A. J., Filippenko, A. V., Chevalier, R. A., Fesen, R. A., Fransson, C., Kirshner,
  R. P., \& Leibundgut, B. 1999a, \pasp, 111, 313
\bibitem[Van Dyk et al.~2000]{van00} Van Dyk, S.~D., Peng, C.~Y., King, J.~Y.,
  Filippenko, A.~V., Treffers, R.~R., Li, W., \& Richmond, M.~W. 2000, PASP,
  112, 1532
\bibitem{}Woosley, S. E., \& Weaver, T. A. 1986, \araa, 24, 205
\bibitem{}Woosley, S. E., \& Weaver, T. A. 1995, \apjs, 101, 181
\bibitem{}Zampieri, L., \etal 2003, \mnras, 338, 711
\bibitem{}Zaritsky, D., Hill, J.~M., \& Elston, R. 1990, \aj, 99, 1108
\bibitem[Zwicky 1964]{zwi64} Zwicky, F. 1964, \apj, 139, 514
\end{thebibliography}
\end{document}